# Silicon nitride grating based planar spectral splitting concentrator for NIR light Harvesting

Ameen E[1,2], Mohammad H. Tahersima[3], Surendra Gupta[1,2], Volker J. Sorger[3] and Bala Pesala[1,2*]

[1]Academy of Scientific and Innovative Research, Ghaziabad, India
[2]CSIR-Central Electronics Engineering Research Institute Chennai, India
[3]George Washington University, Washington DC, USA
*e-mail: balapesala@gmail.com

*Abstract—* We design a multi-layered solar spectral splitting planar concentrator for near infrared (NIR) light energy harvesting application. Each layer includes a silicon nitride based subwavelength diffraction grating on top of a glass substrate that is optimized to diffract the incoming solar radiation in a specific band from a broad spectral band (700-1400 nm in the NIR region) into guided modes propagating inside the glass substrate. The steep diffraction angle due to subwavelength grating results in concentrated light at the edge of each layer where it is then converted to electricity using a photovoltaic cell. The spectral splitting planar concentrator shows an overall NIR guiding efficiency of ~18%, and power conversion efficiency of ~11%. The design can be potentially used for building integrated photovoltaics application.

## 1. Introduction

Most of the energy in solar spectrum (AM1.5) lies in a broad spectral range from ~300 nm to ~1700 nm, spanning the ultraviolet to the infrared (>90%). The discrepancy between narrow wavelength band of a semiconductor energy gap and the broadband solar spectrum limits the energy conversion efficiency. The maximum conversion efficiency of a single junction photovoltaic cell is given by Shockley-Queisser limit [1]. In a photovoltaic cell, when a photon is incident with an energy higher than the band gap of the photovoltaic cell, an electron-hole pair is generated which relaxes to the band edge and the excess energy will generate phonons [2], which results in thermalization losses. In addition, single junction photovoltaic cells also have significant sub-band gap losses. For example, photons with energies lower than the bandgap of the photovoltaic material of the photovoltaic cell do not excite an electron in the valance band of the photovoltaic material to the conduction band and as a result do not contribute to electricity generation [1]. In order to overcome the aforementioned fundamental limitations, multijunction photovoltaic cells are used [3] with high conversion efficiencies, that may exceed 45% [4]. The multijunction photovoltaic cells often include three or more semiconductor junctions that collectively provide multiple band gaps for efficient photovoltaic power generation, each of the one or more junction converts photons from a narrow spectral band of the solar spectrum to electricity. When combined together, the multijunction photovoltaic cells can cover a significant portion of the solar spectrum [5]. Although multijunction photovoltaic cells have high conversion efficiencies they are however prevented from entering the commercial market due to various limitations such as higher cost per area due to higher engineering costs (e.g., lattice matching, current matching, transparency loss, to name a few).

Multijunction photovoltaic cells, however, are not the only approach to capture a broader spectrum of the sun light. There are a variety of alternatives such as: up-down conversion [6], spectral splitting devices [7,8], a combination of algae and photovoltaic cell [9], or a combination of photovoltaic cell and thermo-electric generator [10]. The spectral splitting devices have the potential to capture a significant portion of the solar spectrum by engineering spectral response of the optoelectronic device using subwavelength topological features. These devices can use multiple spectral splitting components and multiple photovoltaic components. Each layer can be independently optimized to obtain maximum electrical output. This eliminates the need for current matching between each layer which is one of the main limitations in multi-junction photovoltaic cells. Spectral splitting can be achieved by numerous methods such as dichroic mirrors [11], luminescent concentrators [12] and diffractive optics [13]. Dichroic mirrors being one of the most explored technique in which two different band gap photovoltaic cells are kept in perpendicular orientation and a dichroic mirror placed at 45° thus reflecting a wavelength band to one photovoltaic cell and transmitting another band to the perpendicular cell [11]. Luminescent based splitter [13, 14] uses a waveguide in combination of a lens with different fluorophores, each one absorbing and emitting specific wavelengths. Luminescent based structures are lossy because they lack the directionality factor i.e. the fluorophores emit in all directions and the band gap of the fluorophores is another limiting factor. Diffraction based spectral splitters uses diffraction structures to split the light [13]. The photovoltaic cells are laterally shifted such that desired wavelength falls on each cell. The diffraction element structures explored so far use reflection, which makes them bulky and limits their concentration ratio. Recently, we have designed a grating based planar light concentrator using SiN grating integrated on a glass substrate for NIR energy harvesting [15]. The gratings help in diffracting the incident light into the guided modes of the glass substrate resulting in solar concentration at the edge where a silicon photovoltaic cell converts the incident light to electricity. The system has the dual advantage of visible light transmission for natural room lighting while harvesting the near infrared light for electricity generation suitable for building integrated photovoltaics. However, the system can only convert a narrow band of NIR spectrum for electricity generation.

Here, we explore a novel silicon nitride subwavelength grating based solar spectral splitting planar concentrator optimized for multiple central wavelengths in the NIR frequency range. The design of spectral splitting planar concentrator can be optimized to capture a significant portion of the solar spectrum covering visible and NIR wavelengths. However, our primary goal in this paper, is to mainly capture the near infrared region of solar spectrum for electricity generation. Each photovoltaic cell is arranged perpendicular to incoming radiation, in contrast to conventional multijunction photovoltaic cells (Fig. 1). Such lateral-to-incident-photon configuration potentially reduces various loss channels, provides high spectral irradiation intensity, reduces fabrication complexity, and hence improves cost-effectiveness. A primary concentrator-like lens is often used in known lateral splitting concentrators [16]. However, lenses introduce system limitations such as angular-dependency and often mandate sophisticated tracking mechanisms. In contrast, our design approach uses simpler broadband diffractive elements to ensure minimal tracking requirements. The spectral splitting approach is significantly simpler than multijunction solar cells in terms of the design methodology and fabrication process. The solar splitting concentrator fabrication process is a three-step process i.e. deposition of the silicon nitride of desired thickness on the glass substrate, lithography and etching. Whereas in the case of multijunction solar cells, starting from material selection till the process of fabrication involves highly sensitive processes. The design process of multijunction solar cells includes various strategies to solve the problems of lattice mismatch, current mismatch, creation of tunnelling junctions and so on. In contrast to this, proposed device can be fabricated separately and then integrated with solar cells at the edges allowing for an additional design degree of freedom.

The grating parameters at each layer must be optimized to guide the solar radiation in the wavelength region corresponding to the bandgap of photovoltaic cell material and to maximize transmission in the rest of the wavelength region. To achieve this spectral selectivity, the grating parameters such as duty cycle, period, and thickness are carefully optimized (Fig. 2). The grating structure is designed to diffract the desired wavelength with diffraction angle higher than the critical angle to excite total internal reflection at the bottom glass-air interface and the glass substrate will act as a waveguide. Since the system is a lossy concentrator due to back coupling at the grating-air interface, the main criteria for optimization is lowering the losses by choosing the best combination of grating parameters. The grating in each Spectral Splitting Converter (SSC) is designed to have maximum guiding efficiency in one of the polarizations (transverse electric). Polarization independent designs can also be designed with slight decrease in overall efficiency [15]. Moreover, the gratings are shallow etched i.e. a layer of the grating material is present beneath grating to increase the angular tolerance of the concentrator.

## 2. Design of the spectral splitting concentrator

One-dimensional grating-based structures are used for each splitting concentrator. The basic device structure consists of a glass substrate with silicon nitride gratings on the top surface and photovoltaic cells attached to the edges of the glass (Fig. 1). The glass substrate attached with photovoltaic cell on both edges is an SSC unit, which are arranged in vertical orientation such that the shorter (longer) wavelength is at the top (bottom) with decreasing bandgap (Fig. 1). These SSC units are also arranged in the order of increasing grating period such that higher wavelengths are not affected and transmitted to the subsequent SSC unit. Each layer of the stack is designed to guide a desired band and transmit the lower energy band of the solar spectrum to the next SSC unit. Each individual layer is optimized to attain higher guiding efficiency corresponding to the attached photovoltaic cell.

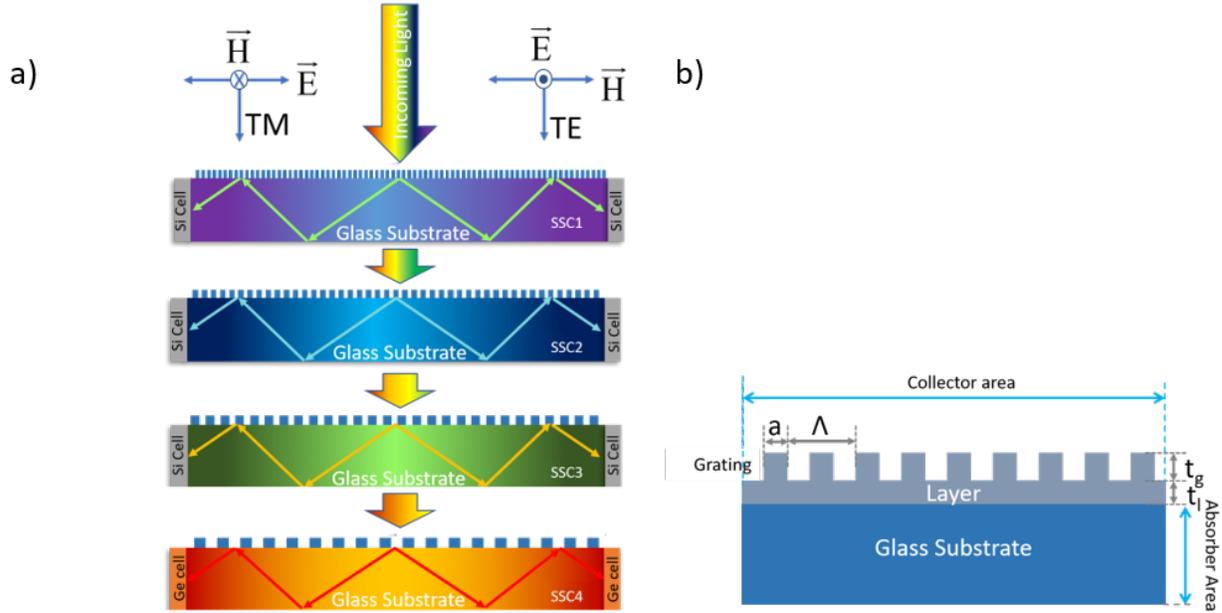

Fig. 1: a) Schematic of solar splitting concentrator with attached photovoltaic cells. SiN diffraction gratings on top of glass substrate act as optical band pass filters for a plurality of wavelengths in the near infrared regime and guide each optical band to their corresponding band-gap selective photovoltaic units. b) Schematic of the grating with the grating parameters annotated. 'Λ' is the grating period, a/Λ is the duty cycle, $t_g$ is the grating thickness and $t_l$ layer thickness. The design optimization is carried out by sweeping of the topological parameters in Finite Difference Time Domain (FDTD) method of the Lumerical software.

The design strategy initiates with the selection of photovoltaic cells to cover the NIR region of the solar spectrum (700-1400 nm) and this spectral range consists of ~40% energy considering AM1.5 incidence. Here, we have chosen Si and Ge photovoltaic cells for the spectral splitting concentrator due to their inherent cost advantages compared to solar cells using III-V materials. The optical guiding efficiency depends on the grating material. Grating semiconductor material is chosen in such a way that it has higher refractive index than the glass substrate, minimum absorption in the NIR wavelengths and high stability in ambient environment conditions. All these conditions are satisfied by silicon nitride. Moreover, silicon nitride is a widely used material in the semiconductor industry, for example, in the photovoltaic cell as anti-reflective coating and in CMOS devices, hence enables a cost-effective and mass-producible material platform.

In order to quantify the performance of this SSCs, we obtain the guiding efficiency, transmission and reflection coefficients from the FDTD simulation. Individual layers are designed and optimized for maximum guiding efficiency in the desired wavelength range and maximum transmission at other wavelengths.

### 3. FDTD simulation results

The design focusses on harvesting the near infrared region of solar spectrum and transmitting the visible light. More than 45% of the solar energy lies in the wavelength range of 700 to 1400 nm [17]. The grating-based concentrator designs do not guide a large wavelength band, the maximum bandwidth achieved in the current design is < 300 nm. This necessitates to design a concentrator with multiple stacks and hence we have chosen 4 stacks to capture a significant portion of solar radiation in this spectral band. Two different band-gap photovoltaic cells are used to extract a significant portion of the near infrared part of solar spectrum from 700 nm to 1400 nm. The Si photovoltaic cells are suitable for energy conversion in lower wavelength band (400 nm to 1000 nm) and Ge for a higher wavelength band (850 nm to 1600 nm). The assumed baseline external quantum efficiencies of the photovoltaic cells are depicted in Fig. 2a, showing a conversion efficiency of more than 80% and 75% for silicon and germanium respectively, in the chosen wavelength region [18]. AM1.5 solar spectrum is shown in figure 2b showing the ultraviolet, visible and the infrared regions.

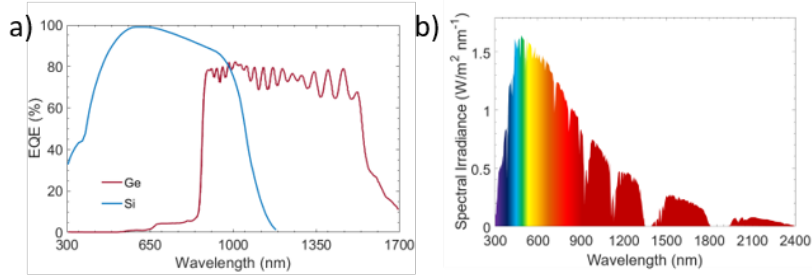

Fig. 2. a) External Quantum Efficiency (EQE) of photovoltaic cells of Ge and Si [15], b) AM1.5 solar spectrum

The grating designs for all the SSC's are shallow etched with a layer same as that of the grating material beneath the gratings. The layer thickness is denoted as $t_l$, the grating thickness denoted by $t_g$, grating duty cycle as DC and grating period as $\Lambda$ (see Fig. 1b). The SSC1 is optimized to have maximum guiding in the wavelength range of 500 nm to 700 nm (Fig. 3a) and maximum transmission at wavelengths higher than 700 nm (Fig. 3b). SSC2 is designed for guiding wavelength in the range of 700 nm to 900 nm (Fig. 3a) and transmit >900 nm (Fig. 3b). We have optimized SSC3 to achieve maximum guiding efficiency at the central wavelength of 923 nm with a bandwidth of 264 nm (Fig. 3a). The SSC4 unit is optimized for Ge photovoltaic cells to guide maximum optical power in the range of 1100 nm - 1300 nm and transmit wavelengths < 1100 nm for visible light transmission. The stack is formed by piling up four individual concentrators consisting of a glass substrate of (2-3 mm thick) each integrated with optimized gratings on top of it. Further, there is a spacing between each of the glass substrates so that each spectral harvester operates independently. The stack is organized in such a way that the energy harvesting happens from lower (higher) to higher (lower) wavelengths (energies) i.e. it is arranged in the ascending order of period from top to bottom. Note that the vertical structure needs to adhere to design criteria so that the lower period gratings do not diffract higher wavelengths which get transmitted to the subsequent section where they are captured. This design methodology ensures that each stack harvests energy from a particular band of wavelength. The guiding efficiency ($\eta$) defined by eqn. 1 is the key metric for optimization.

$$\eta = \int_{700}^{1400} \left( \eta_{G1}(\lambda) + \eta_{G2}(\lambda).T_1(\lambda) + \eta_{G3}(\lambda).T_1(\lambda).T_2(\lambda) + \eta_{G4}(\lambda).T_1(\lambda).T_2(\lambda)).T_3(\lambda) \right) d\lambda \qquad (1)$$

where $\eta_{Gi}$ is the guiding efficiency of $i^{th}$ SSC and $T_i$ is the transmission percentage for $i^{th}$ SSC.

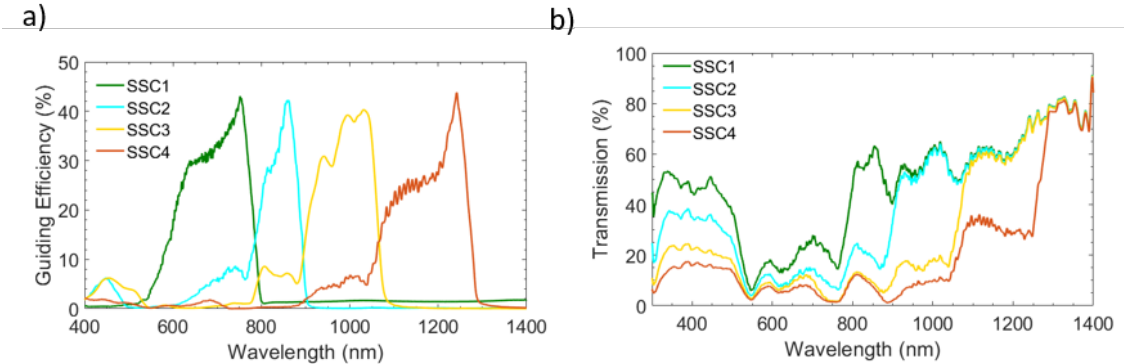

Fig. 3. a) Guiding efficiency of the spectral splitting planar concentrator with period 530 nm (green), period 602 nm (cyan), period 721 (yellow) and period 865 nm (red). b) Transmission efficiency of the spectral splitting planar concentrator with period 530 nm (green), period 602 nm (cyan), period 721 (yellow) and period 865 nm (red).

**Table 1: The optimized grating parameters for each SSC.**

|  | $\Lambda$ (nm) | $t_g$ (nm) | DC | $t_l$ (nm) | Guiding efficiency (%) | Conversion efficiency (%) |
|---|---|---|---|---|---|---|
| SSC1 | 530 | 300 | 0.3 | 50 | 9.69 | 6.21 |
| SSC2 | 602 | 400 | 0.3 | 50 | 4.47 | 2.86 |
| SSC3 | 721 | 400 | 0.4 | 50 | 2.73 | 1.40 |
| SSC4 | 865 | 500 | 0.3 | 50 | 1.52 | 0.80 |

The guiding efficiency for the entire spectral splitting planar concentrator is obtained by the addition of individual guiding efficiency of SSC1 followed by addition of the product of guiding efficiency of the corresponding SSC and transmission percentage of the above SSCs (eqn. 1). Thus, the overall system efficiency depends on three parameters: guiding efficiency, transmission coefficient and IPCE of the photovoltaic cells. The device considers a geometric concentration ratio of 5, defined by the ratio of the absorber area to the collector area. This ensures a reduction in the amount of photovoltaic cell material used and reduces cost and complexities associated in using known multi-junction photovoltaic cells. The optimized grating parameters for all the four SSC's are shown in table 1. The individual guiding (conversion) efficiencies (table 1) for SSC1 is 9.69% (6.21%), for SSC2 it is 4.47% (2.86%), for SSC3 it is 2.73% (1.40%) and that for SSC4 is 1.52% (0.80%). The electrical conversion efficiency $\eta_c$ is calculated using the eqn. 2.

$$\eta_c = \int_{700}^{1400} \eta(\lambda)\, EQE(\lambda)\, FF\, d\lambda \qquad (2)$$

where η is the optical guiding efficiency (eqn. 1), EQE is the external quantum efficiency of the solar cell and FF is the fill factor of the solar cell

Assuming the overall dimension of the glass considered for this calculation is 1 m x 30 mm x 3 mm resulting in an absorber area of 0.03 m² and a collector area of 0.006 m² with a geometric concentration ratio of 5. The combined guiding efficiency for spectral splitting planar concentrator is 18.41% (using eqn. 1) with an electrical conversion efficiency of 11.27% (using eqn. 2). This is almost two times compared to the previous design with a single splitter [15] at the same geometric ratio of 5. The previous design shows guiding efficiency of 9.88% and electrical conversion efficiency of 6.58%. The multijunction solar cells offer record high conversion efficiencies exceeding 45% at a concentration of 690 suns and a temperature of 25 ºC [17]. However, multijunction solar cells cannot be used for the building integrated photovoltaics application due to their high cost and secondly the requirement of higher concentration in order to mitigate the high cost which will in turn demand precise tracking mechanism.

Angular studies for the SSC's are carried out and the same for SSC4 is shown in Fig. 4. The polarization angle for the entire study is transverse electric i.e. parallel to the grating direction. The simulation results show that the angular tolerance for the SSC is ±65° in the parallel direction and ±12° in the perpendicular direction. This ensures that the SSCs are highly angular tolerant structures and hence the device can be implemented in a non-tracking/minimal tracking configuration which avoids the sophisticated opto-mechanical sub systems.

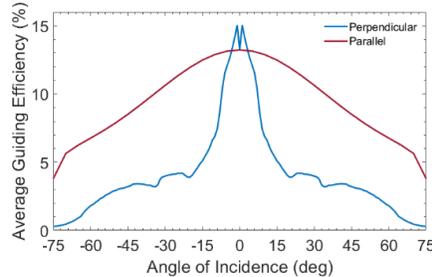

Fig. 4. Angular tolerance of SSC4 with varying the angle of incidence in the perpendicular and parallel directions to the grating bar.

## 4. Conclusion

Germanium and silicon photovoltaic cells are used in conjunction with four different SSC layers to guide the wavelength range of 700 nm to 1400 nm in a descending order. The total optical guiding efficiency is 18.41% for TE incidence which results in an electrical conversion efficiency is 11.27%. Moreover, this eliminates the complexity in design as associated with multijunction solar cells. In future, we will explore polarization independent designs to maximize the conversion efficiencies for both the polarizations. Also we will explore incorporation of thin spiral nanostructured solar cells instead of commercial solar cells [19].

**Disclosures**

The authors declare that there are no conflicts of interest related to this article.